\documentclass[conference]{IEEEtran}
\IEEEoverridecommandlockouts
\usepackage{amssymb}
\usepackage{amsmath,amssymb,amsfonts}
\usepackage{graphicx}
\usepackage{amsmath}  
\usepackage{amssymb}  
\usepackage{geometry}
\usepackage[ruled,linesnumbered]{algorithm2e}

\newgeometry{left =0.673in,right=0.673in, top= 0.75in,bottom=1.1in}

\ifCLASSINFOpdf
\else
\fi

\begin{document}
%
\title{Universal Modem Generation with Inherent Adaptability to Variant Underwater Acoustic Channels: a Data-Driven Perspective}

\author{
    \IEEEauthorblockN{Xiaoquan You, Hengyu Zhang, Xuehan Wang, Jintao Wang}
    \IEEEauthorblockA{Beijing National Research Center for Information Science and Technology (BNRist), \\
    Dept. of Electronic Engineering, Tsinghua University, Beijing, China}
    \IEEEauthorblockA{\{{youxq21}, {zhanghen23}, {wang-xh21}\}@mails.tsinghua.edu.cn, wangjintao@tsinghua.edu.cn} 
}
\maketitle

\begin{abstract}
In underwater acoustic (UWA) communication, orthogonal frequency division multiplexing (OFDM) is commonly employed to mitigate the inter-symbol interference (ISI) caused by delay spread. However, path-specific Doppler effects in UWA channels could result in significant inter-carrier interference (ICI) in the OFDM system. To address this problem, we introduce a multi-resolution convolutional neural network (CNN) named UWAModNet in this paper, designed to optimize the modem structure, specifically modulation and demodulation matrices. Based on a trade-off between the minimum and the average equivalent sub-channel rate, we propose an optimization criterion suitable to evaluate the performance of our learned modem. Additionally, a two-stage training strategy is developed to achieve quasi-optimal results. Simulations indicate that the learned modem outperforms zero-padded OFDM (ZP-OFDM) in terms of equivalent sub-channel rate and bit error rate, even under more severe Doppler effects during testing compared to training.
\end{abstract}
\begin{IEEEkeywords}
deep learning, modem generation, underwater acoustic channels, ZP-OFDM.
\end{IEEEkeywords}


%
\IEEEpeerreviewmaketitle

\section{Introduction}
So far, the concept of the ``Smart Ocean'' has attracted considerable interest, highlighting a promising future for the Internet of Underwater Things (IoUT). IoUT can be applied in various marine applications such as marine resource exploration, defense against enemy attacks, and monitoring underwater pollution, where underwater communication plays a crucial role \cite{9848467,10423132}.

In underwater communication, acoustic waves, due to their lower attenuation in water compared to electromagnetic waves, are capable of achieving long-distance communication and thus have broad development prospects. However, the reflection and refraction of acoustic waves between the surface and seafloor can cause a significant multipath delay spread. Furthermore, the path-specific Doppler effects, due to internal waves, platform, and sea-surface motion, also have a non-negligible impact on underwater acoustic (UWA) communication \cite{10337106,10317502}.

As a result, orthogonal frequency division multiplexing (OFDM) is widely recognized as an effective solution to mitigate the inter-symbol interference (ISI) caused by multipath propagation \cite{7470947}. However, when utilized in underwater acoustic channels, the orthogonality of OFDM subcarriers is compromised by the path-specific Doppler scale, which results in severe inter-carrier interference (ICI) among the subchannels. Additionally, to conserve transmission power allocated to the guard interval, zero-padded OFDM (ZP-OFDM) is preferred in underwater acoustic communication \cite{4554200,7535624}. 

To reduce the impact of ICI, previous research has been devoted to finding mathematical solutions to this problem. Resampling, a traditional method, is based on an assumption that all the paths have the same Doppler scaling factor \cite{4554200}. Some focus on identifying new waveforms suitable for communication over UWA channels, like VBMC waveform \cite{9833981}, but lack mathematical derivation. Furthermore, deep learning-based methods have been proposed to optimize modem structures, specifically modulation and demodulation matrices, in high-mobility scenarios, among which ModNet significantly outperforms traditional OFDM techniques \cite{zhang2024modem}. However, the signal model built in ModNet does not account for fractional delays, limiting its applicability.

To solve these problems, we aim to optimize the modem structure and build a transmission model for UWA communication. Based on a trade-off between the minimum and the average equivalent sub-channel rate, we propose an optimization criterion suitable for deep learning. Following \cite{zhang2024modem,9149229}, we introduce a multi-resolution convolutional neural network (CNN) to enhance the performance of modem generation across UWA channels. Additionally, a two-stage training strategy is proposed to achieve quasi-optimal results. Finally, simulations on equivalent sub-channel rate and bit error rate show that the learned modem outperforms ZP-OFDM, even under more severe Doppler effects during testing compared to training.

$\textit{Notations:}$ Matrices are denoted by bold uppercase letters, vectors by bold lowercase letters, and scalars by normal font. The notation $(\cdot)^H$ indicates the Hermitian transpose, while $\mathbb{E}[\cdot]$ denotes the mathematical expectation.

\section{System Model}
In this section, we present our signal model of the general modem for UWA channels, followed by the modem structure of ZP-OFDM. After that, we introduce the proposal of the evaluation criterion and optimization problem.

\subsection{Signal model and UWA channel}
We denote the data vector as $\mathbf{s}\in\mathbb{C}^{N\times1}$ with zero mean and an autocorrelation matrix $\mathbb{E}[\mathbf{s}\mathbf{s}^H]=\sigma_s^2\mathbf{I}_N$, where $N$ is the number of subcarriers. The modulated signal $\mathbf{x}\in\mathbb{C}^{M\times1}$ can be written as 
\begin{equation}\label{eq1}
	{\mathbf{x = \Phi s}}
\end{equation}
where $\mathbf{\Phi}\in\mathbb{C}^{M\times N}(N\leq M)$ is the modulation matrix.

We construct the baseband signal $x(t)$, which is band-limited to $-B/2\leq f\leq B/2$, by employing a Nyquist interpolation filter.
\begin{equation}\label{eq2}
	x(t)=\sum_{m=-\infty}^{\infty}x[m]\mathrm{ sinc}(B(t-m/F_s)),t\in\mathbb{R}
\end{equation}
where $x(m/F_s)=x[m]$, assumed that the sampling rate $F_s\geq B$, and
\begin{equation}\label{eq3}
	\operatorname{sinc}(x) \stackrel{\text{def}}{=}
	\left\{
	\begin{array}{cll}
		\frac{\sin(\pi x)}{\pi x} &,& \text{if } x \neq 0, \\
		1 &,& \text{if } x = 0.
	\end{array}
	\right.
\end{equation}
In practice, we typically use a waveform with a finite duration $T$, which can only be approximately bandlimited. Define $M = \lfloor F_sT \rfloor$ as the number of samples taken from $x(t)$ within the time interval $0 \leq t \leq T$. We can reformulate $x(t)$ as follows:
\begin{equation}\label{eq4}
	\tilde{x}(t)=\sum_{m=0}^{M-1}x[m]\mathrm{sinc}(Bt-mB/F_s),0 \leq t \leq T.
\end{equation}
Following \cite{9833981}, we assume that $\varepsilon = \|x(t) - \tilde{x}(t)\|^2 \rightarrow 0$ and thus approximate $x(t)$ by $\tilde{x}(t)$.

The transmitted passband signal can be given by
\begin{equation}\label{eq5}
	\begin{aligned}
		\hat{x}(t)
		&=x(t)\mathrm{e}^{\mathrm{j}2\pi f_c t}\\
		&=\sum_{m=0}^{M-1}x[m]\mathrm{sinc}(Bt-mB/F_s)\mathrm{e}^{\mathrm{j}2\pi f_c t}.
	\end{aligned}
\end{equation}

Following \cite{4554200}, the impulse response of the UWA channel can be determined by
\begin{equation}\label{eq6}
	c(\tau,t)=\sum_{p=1}^P A_p(t)\delta(\tau-\tau_p(t))
\end{equation}
where $P$ denotes the number of propagation paths, $A_p(t)$ is the path amplitude and $\tau_p(t)$ is the time-varying path delay. Since the duration of a transmitted signal is less than the channel stationary time, $A_p(t)$ can be considered as a constant value $A_p$, and $\tau_p(t)$ can be approximated by
\begin{equation}\label{eq6-7}
	\tau_p(t)\approx\tau_p-a_pt
\end{equation}
where $a_p$ denotes the path-specific Doppler scaling factor and $\tau_p$ denotes the constant path delay. As a result, $c(\tau,t)$ can be derived as
\begin{equation}\label{eq7}
	c(\tau,t)=\sum_{p=1}^P A_p\delta(\tau+a_pt-\tau_p).
\end{equation}

The received passband signal is 
\begin{equation}\label{eq8}
	\begin{aligned}
		\hat r(t)&=\hat{x}(t)*c(\tau,t)\\
		&=\sum_{p=1}^PA_px((a_p+1)t-\tau_p)\mathrm{e}^{\mathrm{j}2\pi f_c((a_p+1)t-\tau_p)}
	\end{aligned}
\end{equation}
where the impact of noise is neglected for simplicity in illustration.

The baseband signal $r(t)$ can then be derived as
\begin{equation}\label{eq9}
	\begin{aligned}
		r(t)&=\hat r(t)\mathrm{e}^{-\mathrm{j}2\pi f_c t}\\
		&=\sum_{p=1}^P A_p x((a_p+1)t-\tau_p)\mathrm{e}^{\mathrm{j}2\pi f_c(a_pt-\tau_p)}\\
		&=\sum_{p=1}^P\Biggl\{  A_p\left[\sum_{m=0}^{M-1}x[m]\mathrm{sinc}(B((a_p+1)t-\tau_p)-mB/F_s)\right]\\ 
		&\qquad \qquad \times \mathrm{e}^{\mathrm{j}2\pi f_c(a_pt-\tau_p)}\Biggr\}
	\end{aligned}
\end{equation}
with a duration of $0\leq t\leq T+ T_g$ due to the delay spread in the channel. The guard interval $T_g$ is given by 
$T_g= [\tau_p/(a_p+1)]_{\max}$ to prevent interference among different symbols.

Time domain samples $r[m']$ can then be obtained by sampling $r(t)$ at a rate $F_s$, where we have
\begin{equation}\label{eq10}
	\begin{aligned}
		r[m']=\sum_{m=0}^{M-1}x[m]&\Biggl\{ \sum_{p=1}^P \Biggl[A_p \mathrm{e}^{-\mathrm{j}2\pi f_c\tau_p} \times \mathrm{e}^{\mathrm{j}2\pi f_c a_p m'/F_s}\\
		\times &\mathrm{sinc}(B((a_p+1)m'/F_s-\tau_p)-mB/F_s)\Biggr]\Biggr\}
	\end{aligned}
\end{equation}
for $m'=0,1,\cdots,M'-1$, where $M'$ denotes the number of signal samples of $r(t)$, which is set as $\left \lfloor F_s T+F_s T_g\right \rfloor$ to capture all useful signals to maximize information retention.

The received vector $\mathbf{r}\in \mathbb{C}^{M'\times 1}$ can be written in the matrix-vector notation as follows, including the additive white Gaussian noise $\mathbf{w}\sim \mathcal{CN}(0,\sigma_n^2\mathbf{I}_{M'})$.
\begin{equation}\label{eq11}
	\mathbf{r=Hx+w}.
\end{equation}

$\mathbf{H}\in\mathbb{C}^{M'\times M}$ denotes the channel matrix, whose $(m',m)\mathrm{th}$ entry is
\begin{equation}\label{eq12}
	\begin{aligned}
		[\mathbf{H}]_{m',m}&= \sum_{p=1}^P \Biggl[A_p  \mathrm{e}^{-\mathrm{j}2\pi f_c\tau_p} \times \mathrm{e}^{\mathrm{j}2\pi f_c a_p m'/F_s}\\
		&\times \mathrm{sinc}(B((a_p+1)m'/F_s-\tau_p)-mB/F_s)\Biggr].
	\end{aligned}
\end{equation}

To simplify the expression, the channel matrix $\mathbf{H}$ can be expressed in another form as
\begin{equation}\label{eq13}
	\mathbf{H}=\sum_{p=1}^P \xi_p\mathbf{\Lambda}_p\mathbf{\Gamma}_p.
\end{equation}
The complex path gain for the $p\mathrm{th}$ path $\xi_p\in \mathbb{C}$ is
\begin{equation}\label{eq14}
	\xi_p=A_p \mathrm{e}^{-\mathrm{j}2\pi f_c\tau_p}.
\end{equation}
$\mathbf{\Lambda}_p\in\mathbb{C}^{M'\times M'}$ is a diagonal matrix with an $(m',m')\mathrm{th}$ entry as 
\begin{equation}\label{eq15}
	[\mathbf{\Lambda}_p]_{m',m'}=\mathrm{e}^{\mathrm{j}2\pi f_ca_pm'/F_s}.
\end{equation}
The matrix $\mathbf{\Gamma}_p\in\mathbb{C}^{M'\times M}$ has an $(m',m)\mathrm{th}$ entry as 
\begin{equation}\label{eq16}
	[\mathbf{\Gamma}_p]_{m',m}=\mathrm{sinc}(B\gamma_{m'}^{(p)}-mB/F_s).
\end{equation}
Here we adopt the notation as $\gamma_{m'}^{(p)}=(a_p+1)m'/F_s-\tau_p$, noting that when $\gamma_{m'}^{(p)}<0$ or  $\gamma_{m'}^{(p)}>T$, we have 
\begin{equation}\label{eq17}
	[\mathbf{\Gamma}_p]_{m',m}=0.
\end{equation}

The signal after demodulation $\mathbf{y}\in\mathbb{C}^{N\times 1}$ at the receiver can be expressed as
\begin{equation}\label{eq18}
	\mathbf{y}=\mathbf{\Psi}^H\mathbf{r}=\mathbf{\Psi}^H\mathbf{H\Phi s}+\mathbf{\Psi}^H\mathbf{w}=\mathbf{H}_e\mathbf{s}+\mathbf{\Psi}^H\mathbf{w}
\end{equation}
where $\mathbf{H}_e=\mathbf{\Psi}^H\mathbf{H\Phi}$ denotes the equivalent channel and $\mathbf{\Psi}^H\in \mathbb{C}^{N\times M'}$ represents the demodulation matrix, which converts symbols into original dimensions.

\subsection{ZP-OFDM modem structure}
Following \cite{4554200,5352256}, in the ZP-OFDM system, due to the presence of null subcarriers, the modulation matrix $\mathbf{\Phi}_{\mathrm{OFDM}}\in\mathbb{C}^{M\times N}$ typically has fewer columns than rows, which can be split into two parts
\begin{equation}\label{eq19}
	\mathbf{\Phi}_{\mathrm{OFDM}}=\mathbf{F}_M^H\mathbf{X}.
\end{equation}
$\mathbf{F}_M$ denotes the discrete Fourier transform (DFT) matrix with the $(i,j)$th entry
\begin{equation}\label{eq20}
	[\mathbf{F}_M]_{i,j}=\frac{1}{\sqrt{M}}\mathrm{e}^{-\mathrm{j}2\pi (i-1)(j-1)/M}.
\end{equation}
$\mathbf{X}\in\mathbb{C}^{M\times N}$ represents the matrix to extract $N$ columns from $\mathbf{F}_M^H$, ensuring the selected subcarriers are as evenly distributed as possible within the passband.

The demodulation matrix $\mathbf{\Psi}^H\in\mathbb{C}^{N\times M'}$ can be written as
\begin{equation}\label{eq21}
	\mathbf{\Psi}_{\mathrm{OFDM}}^H=\mathbf{X}^H\mathbf{F}_M\mathbf{R}
\end{equation}
where $\mathbf{R}\in \mathbb{C}^{M\times M'}$ represents the matrix to append the last $L=M'-M$ columns of $\mathbf{X}^H\mathbf{F}_M$ on its front.
\begin{equation}\label{eq22}
	\mathbf{R}=
	\begin{bmatrix}
		\mathbf{0}_{(M-L)\times L} & \mathbf{I}_{M-L} & \mathbf{0}_{(M-L)\times L}\\
		\mathbf{I}_L & \mathbf{0}_{L\times (M-L)} & \mathbf{I}_L
	\end{bmatrix}_{M\times M'}.
\end{equation}

\subsection{Proposed criterion and optimization problem}
We aim to identify the modem structure (i.e. modulation matrix $\mathbf{\Phi}$ and demodulation matrix $\mathbf{\Psi}^H$), which will optimize system performance across various channel matrices $\mathbf{H}$. Therefore, we need to establish an appropriate evaluation criterion to model the optimization problem, where the equivalent sub-channel rate proves to be useful, following \cite{zhang2024modem}.

The signal after demodulation $\mathbf{y}$ can be segmented into $N$ sub-channels, each subject to interference and noise. The output of the $n\mathrm{th}$ sub-channel, denoted by $\mathbf{y}[n]$, can be expressed as
\begin{equation}\label{eq23}
	\begin{aligned}
		\mathbf{y}[n]&=[\mathbf{H}_{e}]_{n, n} \mathbf{s}[n]+\sum_{k=0, k \neq n}^{N-1}[\mathbf{H}_{e}]_{n, k} \mathbf{s}[k]\\
		&+\sum_{m'=0}^{M'-1}[\mathbf{\Psi}^{H}]_{n, m'} \mathbf{w}[m']
	\end{aligned}	
\end{equation}
where the last two terms of the sum represent interference and noise, respectively.

Since we find the worst sub-channel has a more significant impact on the transmission, the criterion function is defined as
\begin{equation}\label{eq24}
	f(\mathbf{H}_e)=\sum_{n=0}^{N-1}r_n(\mathbf{H}_e)+KN\cdot \min_n r_n(\mathbf{H}_e)
\end{equation}
where $K\geq 1$ denotes the amplification factor of the worst sub-channel rate.
\begin{footnotesize}
	\begin{flalign}\label{eq25}
		r_n(\mathbf{H}_e)=\log_{2}\left(1+\frac{\left|[\mathbf{H}_{e}]_{n, n}\right|^2}{\sum_{\substack{k=0,\\ k \neq n}}^{N-1}\left|[\mathbf{H}_{e}]_{n, k}\right|^2+\frac{\sigma_n^2}{\sigma_s^2}\sum_{m'=0}^{M'-1}\left|[\mathbf{\Psi}^{H}]_{n, m'}\right|^2}\right)
	\end{flalign}
\end{footnotesize}where $r_n$ denotes the $n$th equivalent sub-channel rate, which is related to the signal-to-noise ratio (SNR) $\sigma_s^2/\sigma_n^2$. Hence, the modulation design can be formulated as an optimization problem
\begin{equation}\label{eq26}
	\max _{\mathbf{\Psi}, \mathbf{\Phi}} \mathbb{E}_{\mathbf{H}}\left[f\left(\mathbf{H}_{e}\right)\right], \mathbf{\Phi} \in \mathbb{C}^{M \times N}, \mathbf{\Psi}^{H} \in \mathbb{C}^{N \times M'}
\end{equation}
Where $\mathbb{E}_{\mathbf{H}}[\cdot]$ represents the expectation conditioned on the distribution of $\mathbf{H}$, we introduce deep learning-based methods to approximate the quasi-optimal solution.

\section{Design of UWAModNet and Training Strategy}

In this section, a network named UWAModNet is introduced to solve the optimization problem. Following \cite{zhang2024modem}, we employ a two-stage training strategy to standardize the modem structure, ensuring compatibility across various channels.
\subsection{UWAModNet structure}
\begin{figure}[!t]
	\centering
	\includegraphics[width=0.80\linewidth]{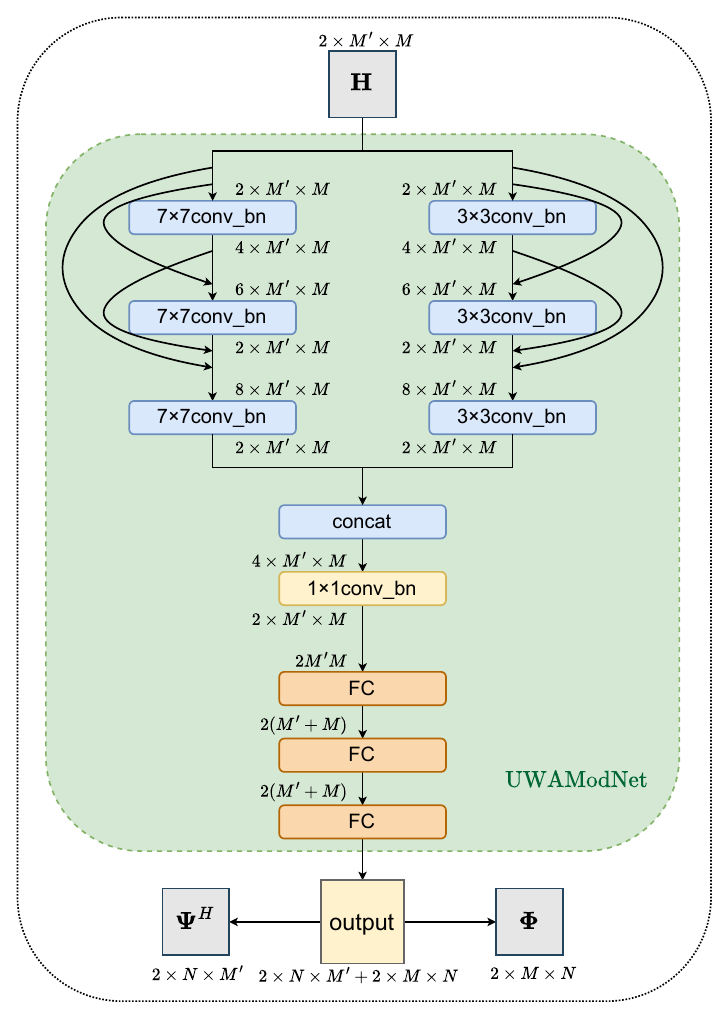}
	\caption{The structure of the proposed UWAModNet. For simplicity, the activation functions and reshape blocks are omitted from the diagram.}
	\label{fig_1}
\end{figure}

As demonstrated in Fig. \ref{fig_1}, we propose a multi-resolution network since the energy distribution of $\mathbf{H}$ exhibits significant localization characteristics. In regions where the energy distribution is more concentrated, smaller convolutional kernels are better able to extract finer features, while in regions where the energy distribution is not concentrated, larger convolutional kernels are adequate. The UWA channel matrix $\mathbf{H}$ is processed as an input image with dimensions $2\times M'\times M$. This image comprises two channels, representing the real and imaginary parts of $\mathbf{H}$. The input image is then processed through two parallel pathways. The first pathway consists of three sequential $7\times 7$ convolutional layers that generate a high-resolution view. In contrast, the second pathway includes three consecutive $3\times 3$ convolutional layers, resulting in a lower resolution. Each convolution is followed by a batch normalization, and a dense connection is employed among the convolution layers, alleviating the over-fitting problem. The outputs from both pathways are then concatenated and integrated using a $1\times 1$ convolutional layer. Subsequently, three fully connected (FC) layers are incorporated to adjust the output size, which is finally divided into $\mathbf{\Phi}$ and $\mathbf{\Psi}^H$. The energies of $\mathbf{\Phi}$ and $\mathbf{\Psi}^H$ are normalized to $N$ and $NM'/M$, respectively, aligning with the ZP-OFDM standards.

It is worth noting that each layer, except for the final fully connected (FC) layer, is followed by a leaky ReLU activation layer to introduce non-linearity. Leaky ReLU is defined as
\begin{equation}\label{eq27}
	\mathrm{LeakyReLU}(x)=
	\left\{
		\begin{array}{ccc}
			x&,&x\geq 0\\
			\beta x&,&x<0\\
		\end{array}
	\right.
\end{equation}
where the negative slope, denoted as $\beta$, is set to a default value of 0.3.

\subsection{Two-stage training strategy}
Within the specified range of channel parameters, our goal is to obtain definite modulation matrix $\mathbf{\Phi}$ and demodulation matrix $\mathbf{\Psi}^{H}$. However, when we simultaneously aim to optimize system performance and ensure the convergence of system outputs, the training outcomes are not entirely satisfactory. Thus, we have revised our training strategy: In Stage I, we focus on enhancing system performance, and in Stage II, we address both system performance and the convergence of system outputs, as is developed in \textbf{Algorithm \ref{algorithm2}} where $E_1$ and $E_2$ denote the training epochs of Stage I and Stage II respectively.

\begin{algorithm}[!t]
	\caption{Training Strategy of the Proposed UWAModNet}
	\label{algorithm2}
	\KwIn{the training dataset $\{\mathbf{H},\mathbf{H}_{e,\mathrm{OFDM}}\}$ and the validation dataset $\{\mathbf{H},\mathbf{H}_{e,\mathrm{OFDM}}\}$ }
	\KwOut{the learned modulation and demodulation matrices $\{\mathbf{\Phi},\mathbf{\Psi}^{H}\}$}
	\BlankLine
	\% \textit{Stage I}
	
	initialize model parameters $\boldsymbol{\theta}$ randomly
	
	\For{$\mathrm{epoch}=1,2,\cdots,E_1$}{
		sample a batch of $\{\mathbf{H},\mathbf{H}_{e,\mathrm{OFDM}}\}$ from the training dataset
		
		compute the output $\{\mathbf{\Phi},\mathbf{\Psi}^{H}\}$ of UWAModNet
		
		compute $\mathrm{loss}_1$ according to equation (\ref{eq24}), (\ref{eq25}) and (\ref{eq28})
		
		validate the performance of the learned modem and update $\boldsymbol{\theta}$
	}
	\% \textit{Stage II}
	
	\For{$\mathrm{epoch}=1,2,\cdots,E_2$}{
		sample a batch of $\{\mathbf{H}_1,\mathbf{H}_{1e,\mathrm{OFDM}}\}$ and another batch of $\{\mathbf{H}_2,\mathbf{H}_{2e,\mathrm{OFDM}}\}$ from the training dataset
		
		compute the corresponding output $\{\mathbf{\Phi}_1,\mathbf{\Psi}_1^{H}\}$ and $\{\mathbf{\Phi}_2,\mathbf{\Psi}_2^{H}\}$ of UWAModNet respectively
		
		compute $\mathrm{loss}_2$ according to equation (\ref{eq24}), (\ref{eq25}) and (\ref{eq29})
		
		validate the performance of the learned modem and update $\boldsymbol{\theta}$
	}
	\% \textit{Generate the final modem}
	
	compute the outputs of UWAModNet with optimized model parameters on the validation dataset
	
	Aggregate the multiple outputs using the average method to obtain the final modem $\{\mathbf{\Phi},\mathbf{\Psi}^{H}\}$
\end{algorithm}

\subsubsection{\textit{\textbf{Stage I: Optimization stage}}}
Channel matrices are selected from the training dataset to update weights of UWAModNet. Based on equation (\ref{eq24}), we define the loss function for this stage as follows.
\begin{equation}\label{eq28}
	\mathrm{loss_1}=f(\mathbf{H}_{e,\mathrm{OFDM}})-f(\mathbf{H}_{e})
\end{equation}
where $\mathbf{H}_{e,\mathrm{OFDM}}=\mathbf{\Psi}^H_{\mathrm{OFDM}}\mathbf{H}\mathbf{\Phi_{\mathrm{OFDM}}}$ denotes the equivalent channel matrix with ZP-OFDM modulation and demodulation matrices. When the value of $\mathrm{loss_1}$ is less than zero, it indicates that the performance of the learned modem surpasses ZP-OFDM.

\subsubsection{\textit{\textbf{Stage II: Convergence stage}}}
The weights derived from Stage I initialize UWAModNet for Stage II. Two different channel matrices $\mathbf{H}_1,\mathbf{H}_2$ selected from the training dataset are fed into our network. Adjustments to the weights will be made according to the output matrices $\{\mathbf{\Phi}_1,\mathbf{\Psi}_1^{H}\},\{\mathbf{\Phi}_2,\mathbf{\Psi}_2^{H}\}$ to ensure convergence. The loss function for this stage is as follows.
\begin{equation}\label{eq29}
	\begin{aligned}			
		\mathrm{loss_2}&=\\
		&\alpha\cdot[f(\mathbf{H}_{1e,\mathrm{OFDM}})-f(\mathbf{H}_{1e})+f(\mathbf{H}_{2e,\mathrm{OFDM}})-f(\mathbf{H}_{2e})]\\
		&+(1-\alpha)\cdot [g(\mathbf{\Phi}_1,\mathbf{\Phi}_2)+g(\mathbf{\Psi}^H_1,\mathbf{\Psi}^H_2)]\\
	\end{aligned}
\end{equation}
where $g(\mathbf{X}_1,\mathbf{X}_2)=\|\mathbf{X}_1-\mathbf{X}_2\|_F$  represents the Frobenius norm and the balance parameter $\alpha$ is defaulted to 0.01. In equation (\ref{eq29}), the first term assesses system performance like $\mathrm{loss_1}$, and the second term quantifies the variance among output modulation and demodulation matrices generated from different channel matrices.

After Stage II, the outputs for various channel matrices largely converge within the specified range of channel parameters. The channel matrices from the validation set are then fed into the well-trained UWAModNet, and the resulting matrices are averaged and normalized to produce the final modulation and demodulation matrices, namely the learned modem.

\section{Simulation Results and Analysis}

\subsection{Experiment settings}
In this section, we describe the UWA channel parameters employed to create the training and validation datasets, some of which are displayed in Table \ref{table}, partially following \cite{9833981}.

\begin{table}[h]
	\caption{Channel Parameters\label{table}}
	\renewcommand{\arraystretch}{1.1}
	\centering
	\setlength{\tabcolsep}{5mm}{
		\begin{tabular}{ll}
			\hline
			Parameter             		& Value      	\\ 
			\hline
			Carrier frequency ($f_c$)   & 15kHz       	\\
			Bandwidth ($B$)   			& 10kHz      	\\
			Sampling rate ($F_s$)       & 10kHz    		\\
			Number of subcarriers ($N$) & 70        	\\
			Symbol duration ($T$) 		& 12.8ms 		\\
			Guard interval ($T_g$) 		& 10.0ms		\\
			Maximum path delay ($\tau_{\max}$)           & 10.0ms   \\
			Maximum Doppler scaling factor ($a_{\max}$)  & 0.001    \\
			Number of paths ($P$)       & 20          	\\ 
			Constellation mapping   	& QPSK			\\
			\hline
	\end{tabular}}
\end{table}

\begin{algorithm}[!t]
	\caption{$\{\mathbf{H}, \mathbf{H}_{e,\mathrm{OFDM}}\}$ Pair Generation Process}
	\label{algorithm1}
	\KwIn{$f_c$, $B$, $F_s$, $N$, $T$, $T_g$, $\tau_{\max}$, $a_{\max}$ and $P$}
	\KwOut{a pair of $\{\mathbf{H}, \mathbf{H}_{e,\mathrm{OFDM}}\}$}  
	\BlankLine
	initialize $\{A_p\}$, $\{\tau_p\}$, $\{a_p\}$, for $p=1,2,\cdots,P$ randomly according to their respective distributions
	
	$M=\left \lfloor F_sT\right \rfloor$, $M'=\left \lfloor F_sT+F_sT_g\right \rfloor$
	
	initialize $\mathbf{H}=\mathbf{0}_{M'\times M}$
	
	\For{$p=1,2,\cdots,P$}{
		compute $\xi_p$ according to equation (\ref{eq14}) 
		
		compute $\mathbf{\Lambda}_p$ according to equation (\ref{eq15}) 
		
		compute $\mathbf{\Gamma}_p$ according to equation (\ref{eq16}) and (\ref{eq17})
		
		$\mathbf{H}\leftarrow \mathbf{H} + \xi_p\mathbf{\Lambda}_p\mathbf{\Gamma}_p$
	}
	
	compute $\mathbf{F}_M$ according to equation (\ref{eq20})
	
	compute the extraction matrix $\mathbf{X}$, ensuring the selected $N$ subcarriers are as evenly distributed as possible within the passband.
	
	compute $\mathbf{\Phi}_{\mathrm{OFDM}}$ according to equation (\ref{eq19})
	
	compute $\mathbf{\Psi}_{\mathrm{OFDM}}^H$ according to equation (\ref{eq21})
	
	compute the equivalent channel matrix for ZP-OFDM system $\mathbf{H}_{e,\mathrm{OFDM}}=\mathbf{\Psi}^H_{\mathrm{OFDM}}\mathbf{H}\mathbf{\Phi_{\mathrm{OFDM}}}$ 
	
	separate the real and imaginary parts from $\mathbf{H}$ and recombine them into a $2\times M'\times M$ tensor
	
	separate the real and imaginary parts from $\mathbf{H}_{e,\mathrm{OFDM}}$ and recombine them into a $2\times N\times N$ tensor
\end{algorithm}

Based on the parameters in Table \ref{table}, we deduce that $M=\left \lfloor F_sT\right \rfloor=128$, $M'=\left \lfloor F_sT+F_sT_g\right \rfloor=228$, and thus the number of null subcarriers $M - N = 58$. In addition, following \cite{9833981}, the path amplitude $A_p$ is distributed as $A_p\stackrel{i.i.d}{\sim}\mathcal{CN}(0,1)$, the path delay $\tau_p$ follows a uniform distribution $\tau_p\stackrel{i.i.d}{\sim}\mathcal{U}(0,\tau_{\max})$, and the path Doppler scaling factor $a_p$ conforms to a uniform distribution $a_p\stackrel{i.i.d}{\sim}\mathcal{U}(1/(1+a_{\max})-1,a_{\max})$ for $p=1,2,\cdots,P$. 

The training dataset consists of 15,000 UWA channel matrices $\mathbf{H}$ and their corresponding equivalent channel matrices for the ZP-OFDM system $\mathbf{H}_{e,\mathrm{OFDM}}$. The validation dataset contains 5,000 pairs of $\{\mathbf{H}, \mathbf{H}_{e,\mathrm{OFDM}}\}$. The training set is used to adjust UWAModNet weights, while the validation set helps to evaluate the performance during training and finalize the learned modem. Additionally, a testing dataset with 10,000 pairs of $\{\mathbf{H}, \mathbf{H}_{e,\mathrm{OFDM}}\}$ is used to generate equivalent sub-channel rate curves and bit error rate curves for system performance evaluation. The process of generating a pair of $\{\mathbf{H}, \mathbf{H}_{e,\mathrm{OFDM}}\}$ is shown in \textbf{Algorithm \ref{algorithm1}}.

For other parameters, the amplification factor $K$ in equation (\ref{eq24}) is set to 10 by default. The Adam optimizer, configured with $\beta_1=0.9,\beta_2=0.999$, learning rate $lr=1\times10^{-3}$, and $\epsilon=1\times 10^{-8}$, is used to train UWAModNet. Furthermore, both Stage I and Stage II of our training are configured with 400 epochs and a batch size of 100, with our UWAModNet operating at a signal-to-noise ratio (SNR) of 20 dB.

\subsection{Performance of the learned modem}
In this section, two figures are presented to demonstrate the performance of the learned modem, one showing the equivalent sub-channel rate and the other depicting the bit error rate.

\begin{figure}[!t]
	\centering
	\includegraphics[width=0.90\linewidth]{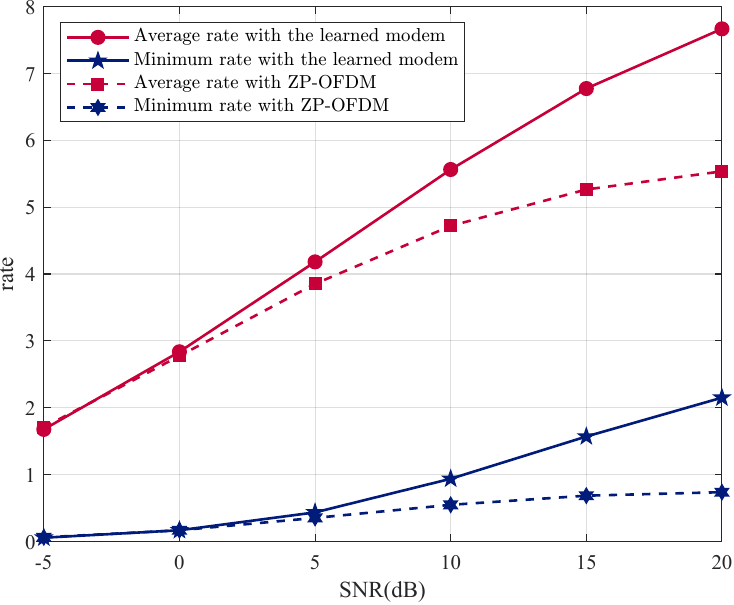}
	\caption{Average and minimum equivalent sub-channel rate with $a_{\max}=0.001$.}
	\label{fig_2}
\end{figure}

Fig. \ref{fig_2} shows how average and minimum equivalent sub-channel rates vary with SNR ranging from -5 to 20 dB with $a_{\max}=0.001$. At an SNR of 20 dB, the learned modem enhances the average equivalent sub-channel rate by 38.5\% and boosts the minimum equivalent sub-channel rate by 190.8\% compared to ZP-OFDM. The figure demonstrates that the learned modem achieves a higher average and minimum rate, indicating significant improvements in overall and worst performance.
\begin{figure}[!t]
	\centering
	\includegraphics[width=0.90\linewidth]{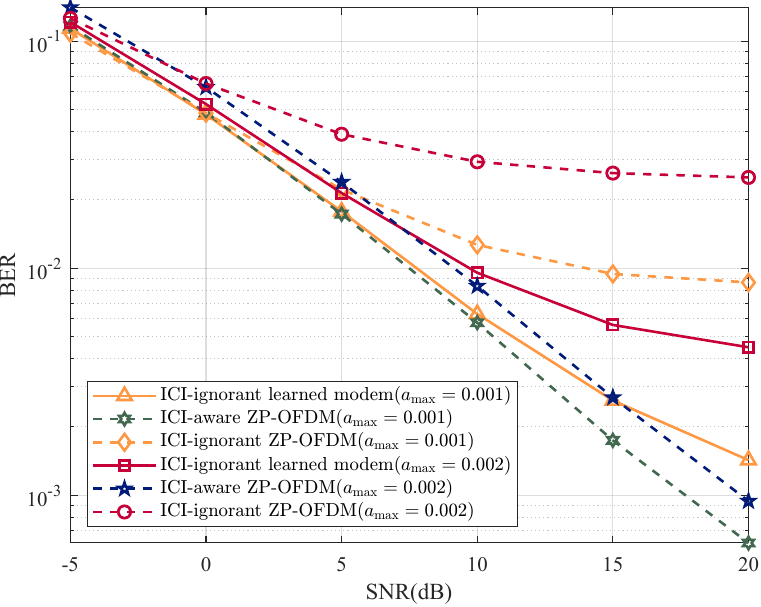}
	\caption{Comparison of bit error rate for UWAModNet and ZP-OFDM under testing constraints $a_{\max}=0.001$ and $a_{\max}=0.002$, originally trained at $a_{\max}=0.001$. }
	\label{fig_3}
\end{figure}

As for the bit error rate, at the transmitter, the bit sequence $\mathbf{b}$ is translated to the symbol sequence $\mathbf{s}$ using QPSK constellation mapping with Gray coding. At the receiver, we perform linear zero-forcing (LZF) equalization on the demodulated signal $\mathbf{y}$ as follows.
\begin{equation}\label{eq30}
	\mathbf{z}=\left(\hat{\mathbf{H}}^H\hat{\mathbf{H}}\right)^{-1}\hat{\mathbf{H}}^H\mathbf{y}
\end{equation}
where $\mathbf{z}$ denotes the output of the equalizer. We show the performance of both ICI-aware LZF equalization ($\hat{\mathbf{H}}=\mathbf{H_e}$) and ICI-ignorant one-tap LZF equalization ($\hat{\mathbf{H}}=\mathrm{diag}(\mathbf{H_e})$) under constraints $a_{\max}=0.001$ and $a_{\max}=0.002$.

In Fig. \ref{fig_3}, the learned modem achieves comparable BER performance to ICI-aware ZP-OFDM but at a lower computational cost, and it significantly outperforms ICI-ignorant ZP-OFDM. At an SNR of 20 dB, with $a_{\max}=0.001$, the learned modem declines BER by 84\% compared to ICI-ignorant ZP-OFDM. 

Under poorer channel conditions, with $a_{\max}=0.002$, the learned modem, originally trained at $a_{\max}=0.001$, suffers some performance loss but still outperforms ICI-ignorant ZP-OFDM, demonstrating good robustness.

\section{Conclusion}
In this paper, to effectively reduce the severe ICI caused by the Doppler scale in the UWA channels, a data-driven modem generation method has been introduced. In order to derive a learned modem within a specified range of UWA channel parameters, we have proposed an optimization criterion suitable for deep learning, developed a multi-resolution CNN named UWAModNet, and made its corresponding two-stage training strategy. We have performed simulations and demonstrated that the learned modem has a better performance than ZP-OFDM. In order to make the scheme more practical, future work may focus on the corresponding channel estimation methods or adjustments to the experiment settings of the dataset generation.


%

\bibliographystyle{IEEEtran}
\bibliography{IEEEabrv,ref}

\end{document}